\documentclass[aps,prl,twocolumn,
]{revtex4}
\usepackage{amsmath}
\usepackage{graphicx}
\usepackage{amsfonts}
\usepackage{amssymb}
\usepackage{amsmath}
\usepackage{bbold}
\usepackage{color}
\usepackage{comment}

\newcommand{\up}{\uparrow}
\newcommand{\down}{\downarrow}
\renewcommand{\k}{{\bf k}}
\newcommand{\eb}{\epsilon_B}
\newcommand{\p}{{\bf p}}
\newcommand{\q}{{\bf q}}
\newcommand{\0}{{\bf 0}}

\newcommand{\ket}[1]{\left|{#1}\right.\rangle}
\newcommand{\ef}{\epsilon_F}
\newcommand{\eq}{\epsilon_{\q}}

\newcommand{\ep}{\epsilon_{\p}}
\newcommand{\ek}{\epsilon_{\k}}
\newcommand{\ekq}{\epsilon_{\q-\k}}
\newcommand{\ekkq}{\epsilon_{\k+\k'-\q}}
\newcommand{\nn}{\nonumber}
\newcommand{\op}{\omega_z}
\newcommand{\T}{{\cal T}}
\newcommand{\F}{{\cal F}}
\newcommand{\n}{\vec n}
\begin{document}

\title{High-polarization limit of the quasi-two-dimensional Fermi
  gas}

\author{Jesper Levinsen}
\email{jfl36@cam.ac.uk}
\affiliation{T.C.M. Group, Cavendish Laboratory, JJ Thomson Avenue, Cambridge, CB3 0HE, United Kingdom} 

\author{Stefan K. Baur}
\affiliation{T.C.M. Group, Cavendish Laboratory, JJ Thomson Avenue, Cambridge, CB3 0HE, United Kingdom} 

\date{\today}

\begin{abstract}



  We demonstrate that the theoretical description of current
  experiments of quasi-2D Fermi gases requires going beyond usual 2D
  theories. We provide such a theory for the highly spin-imbalanced
  quasi-2D Fermi gas. For typical experimental conditions, we find that the
  location of the recently predicted polaron-molecule transition is
  shifted to lower values of the vacuum binding energy due to the
  interplay between transverse confinement and many-body physics. The
  energy of the attractive polaron is calculated in the 2D-3D
  crossover and displays a series of cusps before converging towards
  the 3D limit. The repulsive polaron is shown to be accurately
  described by a 2D theory with a single interaction parameter.

\end{abstract}


\maketitle



Recently, several experimental groups have realized quasi
two-dimensional (2D) Fermi gases with ultra-cold
atoms~\cite{Frohlich:2011cr,Martiyanov:2010kl, Dyke:2011tg,
  Feld:2011nx, sommer2011_2D, Zhang:2012uq, Koehl2012}. A key
question which arises in this context is whether current theories of
the 2D Fermi gas may accurately be applied to model the quasi-2D
experiments?  We demonstrate that this is indeed the case for the
so-called upper branch. On the other hand, a quantitatively accurate
description of the lower branch requires us to take the confinement
explicitly into account as interactions are increased from weak to
strong. As a central result, we show that the puzzling discrepancy between
the location of the polaron-molecule transition in the recent
experiment~\cite{Koehl2012} and the theoretical 2D
prediction~\cite{Parish:2011vn} may be viewed as a consequence of the
quasi-2D nature of the experiment.

Typical experiments with ultra-cold gases achieve lower dimensionality
by confining a three-dimensional (3D) Fermi gas into a single layer
(or a stack of layers) by a tight transverse harmonic oscillator
potential $V(z)=\frac{1}{2} m \omega_z^2 z^2$. If the temperature $T$
and Fermi energy $\ef= \hbar^2 k_F^2/2m$, where $k_F$ is the Fermi
wave vector and $m$ is the atomic mass, satisfy $k_B T \ll \ef \ll
\hbar \omega_z$ the gas is quantum degenerate and collisions can be
considered to be quasi-2D as transverse degrees of motion are frozen
out~\cite{Petrov:2001fk}.

In order to quantify the effects of the quasi-2D nature of current
experiments, we consider the highly population imbalanced limit of a
single spin-$\down$ impurity in a Fermi sea of spin-$\up$ atoms. In
this limit, it was predicted that the Fermi sea can destroy pairing
(if interactions are sufficiently weak) and give rise to a Fermi
polaron
\cite{PhysRevB.77.020408,PhysRevA.74.063628,PhysRevLett.102.230402},
an impurity dressed by excitations of the Fermi sea, in addition to
the molecular state appearing at stronger coupling.
\begin{figure}
\includegraphics[width=\columnwidth]{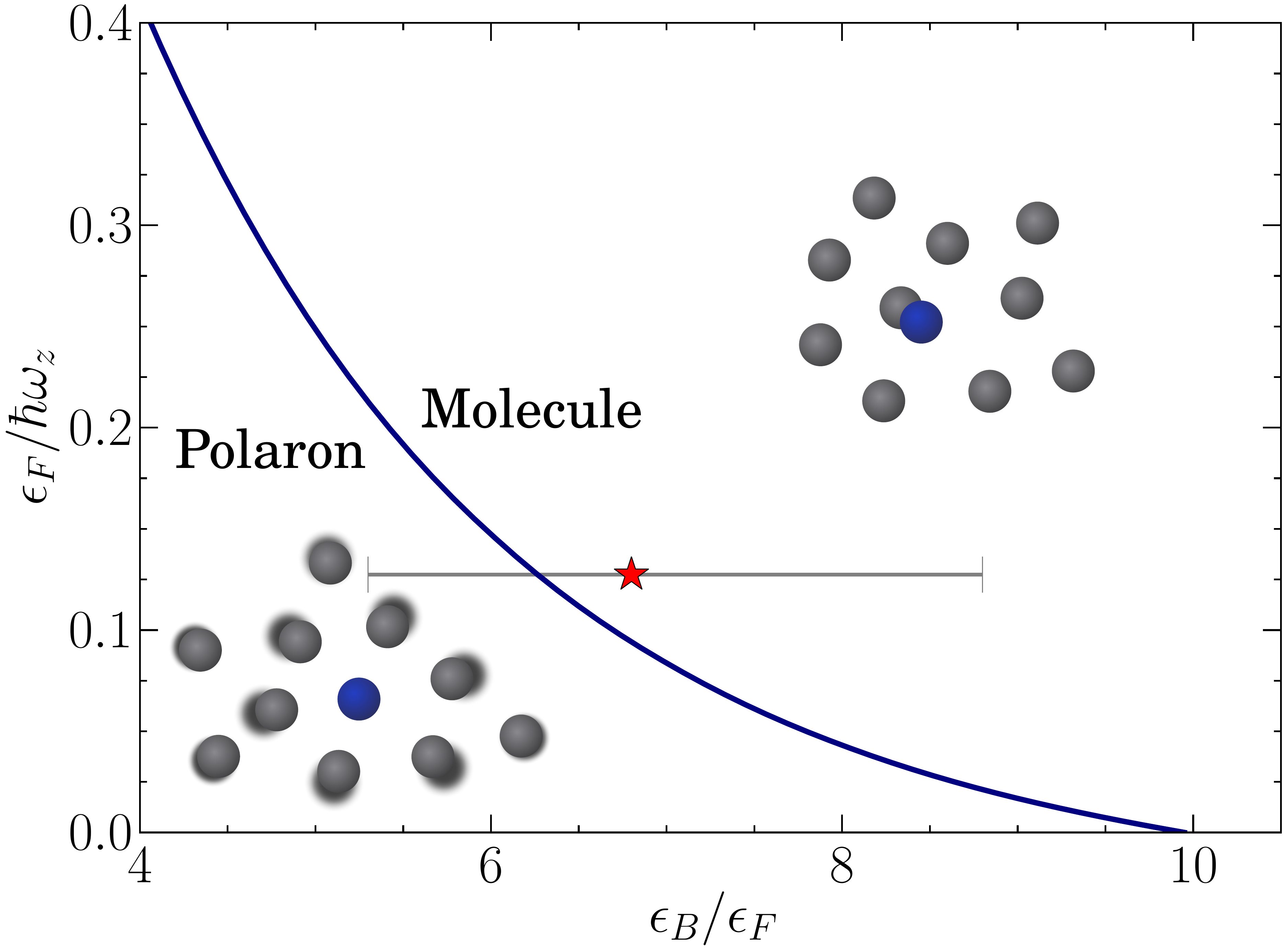}
\caption{(color online). Phase diagram of a highly polarized quasi-two
  dimensional Fermi gas. Below the line, the single spin-$\down$
  impurity exists as a polaron dressed by the Fermi sea of spin-$\up$
  atoms. For large binding energies, the impurity forms a molecular
  bound-state. The location of the transition depends on the strength
  of the external confinement: At higher density (or weaker
  confinement) the transition occurs at lower $\eb/\ef$. The result
  from the recent experiment by Koschorreck {\it et
    al.}~\cite{Koehl2012} is also shown (star).}
\label{fig:phasediagram}
\end{figure}
From a dimensional analysis of the quasi-2D Fermi gas, the location of
the polaron-molecule transition should depend on $\eb/\ef$, where
$\eb>0$ is the vacuum binding energy, and an additional parameter
describing the quasi-2D confinement (i.e. $\ef/\hbar \omega_z$). This
is to be contrasted with the previously studied case of infinitely
tight 2D trapping~\cite{Parish:2011vn}, whose properties depend solely
on the dimensionless interaction parameter $\eb/\ef$. As shown in
Fig. \ref{fig:phasediagram}, we find that the value of $\eb/\ef$ where
this transition occurs strongly depends on the value of $\ef/\hbar
\omega_z$. The transition is shifted to lower $\eb/\ef$ at higher
densities, in good agreement with the experimental measurement of
Ref.~\cite{Koehl2012}.  The 2D limit considered in
Ref.~\cite{Parish:2011vn} is recovered as $\ef/\hbar \omega_z
\rightarrow 0$. The 2D description breaks down as the size of
molecules (or polarons) approaches the harmonic oscillator length of
the transverse confinement $l_z=\sqrt{\hbar/ m \omega_z}$.

We also demonstrate how this quasi-2D physics is manifested in the
energy of the repulsive polaron of the quasi-2D upper branch
gas~\cite{schmidt2011, Ngampruetikorn:2011fk}. As we will argue, in
the repulsive branch, a single effective interaction parameter
describing quasi-2D collisions provides a sufficient description as
long as the (non-universal) lifetime of the repulsive polaron is
large.

Finally, we present results for the energy of an impurity in the
dimensional crossover regime from quasi-2D to 3D, where $\ef\gtrsim
\hbar \omega_z$. This regime was studied experimentally in
Ref.~\cite{Dyke:2011tg}, and also very recently in
Ref.~\cite{Zhang:2012uq} where features in RF-spectra were interpreted
as transitions between polaronic states.  The energy is found to
exhibit an intriguing behavior, as it has a cusp each time the
chemical potential of spin-$\up$ particles admits an extra harmonic
oscillator level. Additionally, the energy is found to quickly
approach the 3D limit.

The low energy scattering of two particles in a two-dimensional
geometry is described through the $s$-wave scattering amplitude
\cite{LL}
\begin{equation}
f(q)=\frac{2\pi}{\ln\left[1/(q a_{2\text{D}})\right]+i\pi/2},
\label{eq:scat2d}
\end{equation}
with $q$ the relative momentum. In the following we work in units
where $\hbar=1$. The scattering at low energies is controlled by the
parameter $a_{2\text D}$ which has the dimension of length. $a_{2\text
  D}$ may be extracted in experiments from the low energy behavior of
the scattering cross section $\sigma=|f(q)|^2/4q.$

Petrov {\em et al.} \cite{Petrov:2001fk} demonstrated how the
scattering amplitude in a quasi-2D geometry is related to
Eq. \eqref{eq:scat2d}: In three dimensions, the properties of an
atomic gas interacting close to a broad Feshbach resonance is
characterized by a single few-body parameter, the scattering length
$a_s$. The energy dependence of the quasi-2D scattering amplitude
$f_{00}$ at energy $\epsilon=k^2/m<\op$ was found to be
\begin{equation}
f_{00}(\epsilon)=\frac{2\sqrt{2\pi}}{l_z/a_s-\F(-\epsilon/\omega_z)},
\label{eq:Petrov}
\end{equation}
where incoming and outgoing particles are assumed to be in the ground
state of the relative motion in the harmonic oscillator potential.
We use the definition of $\F$ \cite{RevModPhys.80.885}
\begin{equation}
\F(x)=\int_0^\infty\frac{du}{\sqrt{4\pi u^3}}\left(1-\frac{e^{-xu}}
{\sqrt{[1-\exp(-2u)]/2u}}\right).
\end{equation}
As opposed to scattering in 3D, the two-dimensional scattering always
admits a bound state, whose binding energy $\eb>0$ satisfies
$l_z/a_s=\F(\eb/\omega_z)$. At low energies, $|\epsilon|\ll
\omega_z$, the function $\F$ takes the form
\begin{equation}
  \F(x)\approx\frac1{\sqrt{2\pi}}\ln\left(\pi x/B \right)+\frac{\ln2}
{\sqrt{2\pi}}x+{\cal O}(x^2)
\label{eq:Gtaylor}
\end{equation}
with $B\approx0.905$ \cite{Petrov:2001fk,RevModPhys.80.885}. Keeping
the first term on the r.h.s., the asymptotic low energy expression
(\ref{eq:scat2d}) is recovered with
\begin{eqnarray}
\label{eq:q2da2d}
a_{2\text{D}}=l_z \sqrt{\pi/B}\exp(-\sqrt{\pi/2}l_z/a_s).
\end{eqnarray}
As demonstrated below, going beyond the leading order in the low
energy expansion of $\F(x)$ is important for the understanding of
current experiments on Fermi gases confined to quasi-2D.

We now turn the problem of a single $\down$ impurity in a Fermi sea of
$\up$ particles. A central tool in the study of many-body problems is
the $T$-matrix, describing the forward scattering between the impurity
and a spin-$\up$ atom at total 2D momentum $\q$ and energy
$\epsilon$. In vacuum, the $T$-matrix is simply related to the
scattering amplitude
\begin{equation}
\T_0(\q,\epsilon)=f_{00}(\epsilon-\eq/2)/m,
\label{eq:t0}
\end{equation}
with the free particle dispersion $\eq=q^2/2m$. The presence of the
Fermi sea couples the center of mass and relative harmonic oscillator
modes and the $T$-matrix depends on harmonic oscillator quantum
numbers $n_1,\,n_2$ and $n_1',\,n_2'$ of incoming and outgoing
particles. The formalism for the $T$-matrix in the presence of the
Fermi sea was recently developed in Ref. \cite{Pietila:2011hc} where
it was demonstrated that the full $T$ matrix, $T^{n_1n_2}_{n_1'n_2'}$,
can be written in terms of a $T$-matrix depending only on center of
mass quantum numbers $N,\,N'$. For details see
Ref. \cite{Pietila:2011hc} and the Supplementary material.

\begin{figure}
\includegraphics[width=.8\columnwidth]{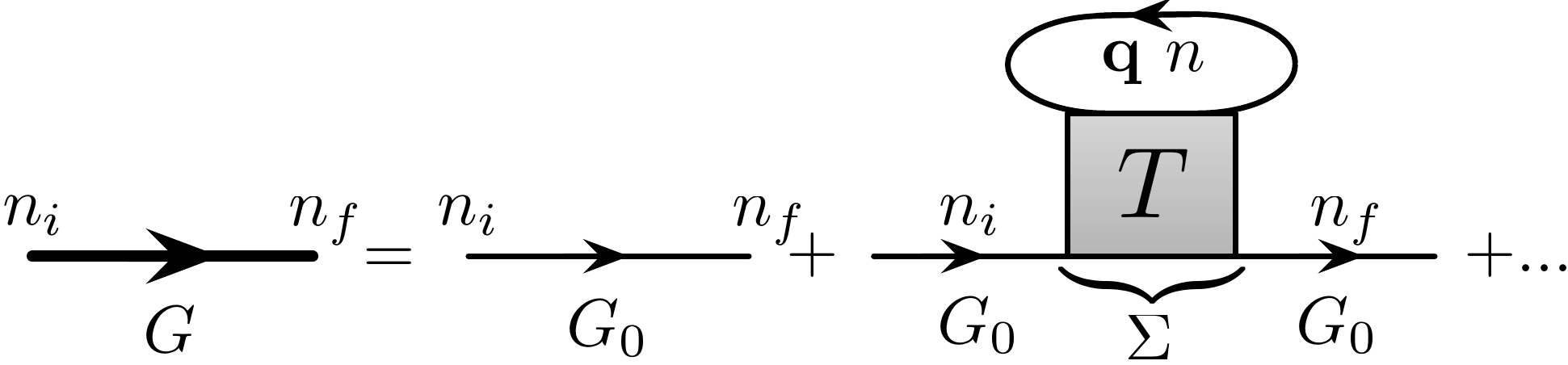}
\caption{The first two terms of the geometric series for the
  propagator of the impurity dressed by a particle-hole excitation.}
\label{fig:feynman}
\end{figure}

In the limit of weak attractive interactions, the impurity exists as a
quasi-particle dressed by a cloud of majority atoms.  This impurity
problem has been widely studied theoretically in both three
\cite{PhysRevA.74.063628,Combescot:2007bh} and two dimensions
\cite{Zollner:2011fk,Parish:2011vn,schmidt2011,Klawunn2011,Ngampruetikorn:2011fk}
using variational and diagrammatic methods. These approaches
correspond to a systematic expansion in the number of particle-hole
pair excitations generated by the interaction between the impurity and
the Fermi sea and provide a strict upper bound on the impurity energy
across the regime of strong interactions. In the limit of weak
interactions they match perturbation theory. In 3D, the accuracy of
this analytic approach has been validated by both exact quantum
Monte-Carlo calculations \cite{PhysRevB.77.020408}, and also by
precise experiments with quantum simulators employing optically
trapped ultra-cold quantum gases~\cite{Kohstall:2011uq}. Indeed, a
systematic expansion in particle-hole pair excitations has
demonstrated an approximate cancellation of contributions from two or
more pairs \cite{PhysRevLett.101.050404}. This result is independent
of dimensionality and explains the remarkable agreement between theory
and experiment. A priori, there is no reason to believe that the
approach is not approximately valid also in quasi-2D, although
corrections to destructive interference are expected to depend on
dimensionality. Here we provide an extension of the diagrammatic
approach to the experimentally relevant quasi-2D setting.

Consider the propagation of the impurity, initially in the state of
harmonic oscillator quantum number $n_i$. Interactions with a
particle from the majority Fermi sea may change the state into a final
state with quantum number $n_f$. Conservation of parity further
restricts $n_i-n_f$ to be even. In the single particle-hole
approximation, the propagator of the impurity then takes the form of a
matrix equation
\begin{equation}
  G(\p,\epsilon)=\left[G_0^{-1}(\p,\epsilon)-\Sigma
    (\p,\epsilon) \right]^{-1},
\label{eq:polaron}
\end{equation}
obtained by summing the geometric series illustrated in
Fig. \ref{fig:feynman}. Note that the diagrams are formally identical
to those considered in 3D \cite{Combescot:2007bh}, however each
propagator is now assigned a harmonic oscillator quantum number which
is summed over intermediate states. The bare propagator is given by
the diagonal matrix $G_0^{n
  n'}(\p,\epsilon)=\delta_{nn'}/\left[\epsilon-\epsilon_{\p
    n}
  +i0\right]$ and the self energy is
\begin{equation}
  \Sigma_{n_1n_2}(\p,\epsilon)=\sum_{\q,n}T_{n_2n}^{n_1n}(\p+\q,\epsilon+
\epsilon_{\q n})\;n_{F\up}(\q,n),
\end{equation}
where $\epsilon_{\p n}\equiv\ep+n\op$.  The Fermi function $n_F$ takes
the value 1 if the state with momentum $\q$ and harmonic oscillator
quantum number $n$ is occupied in the $\up$ Fermi sea, and zero
otherwise. The energy of the polaron corresponds to a pole of
$G(\p,\epsilon)$. From Eq. (\ref{eq:polaron}) the energies of both the
attractive and the repulsive polaron may then be obtained.

\begin{figure}
\includegraphics[width=.99\columnwidth]{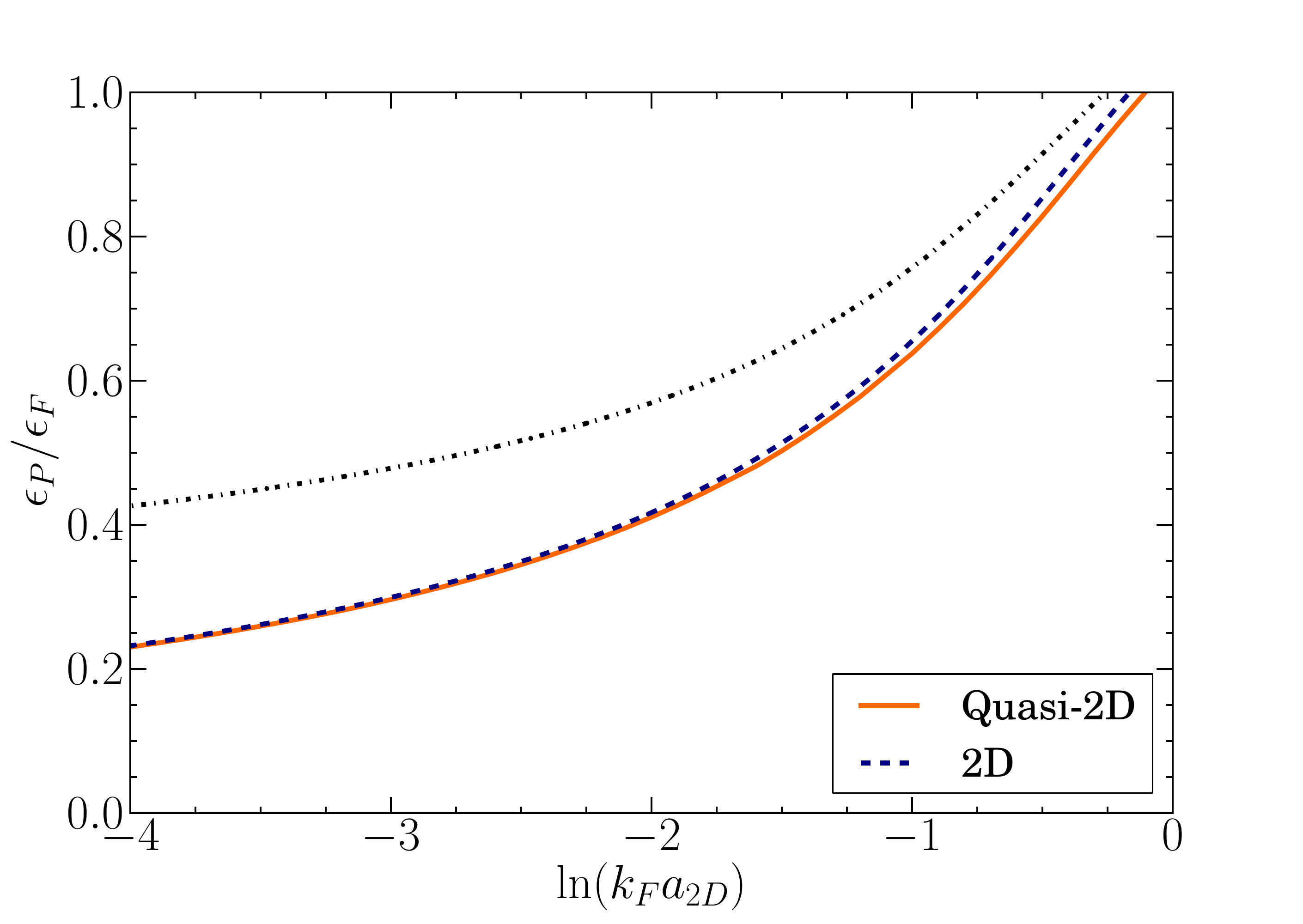}
\caption{\label{fig:rep}
  (color online). Energy $\epsilon_P/\ef$ of an impurity
  interacting repulsively with a Fermi sea as a function of
  interaction strength. The energy of a quasi-2D impurity is shown for
  $\ef/\omega_z=0.1$ (solid line). Additionally we show the 2D results
  of Ref. \cite{Ngampruetikorn:2011fk} using the often used assumption
  $a_{2D}=1/\sqrt{m\eb}$ (dotted), and also using
  Eq. \eqref{eq:scat2d} as definition of $a_{2D}$ (dashed).}
\end{figure}

The energy of the repulsive polaron as a function of interaction
strength is shown in Fig. \ref{fig:rep}. As opposed to the attractive
polaron, our calculation demonstrates that the energy of the repulsive
polaron depends solely on the interaction parameter $k_F a_{2D}$. This
is to be expected as $|\epsilon_P| \ll 
\omega_z$ and thus the energy should match previous studies of the
repulsive polaron in 2D \cite{schmidt2011,
  Ngampruetikorn:2011fk}. Furthermore, Fig. \ref{fig:rep} demonstrates
that in order to compare the predictions of these 2D theories to
experiments with quasi-2D Fermi gases, one should not match the
binding energy of the 2D theory to that of the quasi-2D
system. Instead the 2D scattering length $a_{2D}$ defined in
\eqref{eq:scat2d} and calculated by using the low energy quasi-2D
scattering amplitude in Eqs. (\ref{eq:Petrov})-(\ref{eq:q2da2d})
should be used. Note that while the energy of the repulsive polaron
depends solely on $k_Fa_{2D}$, the decay rate depends sensitively on
the energy difference between the repulsive polaron and the lower
lying states \cite{Massignan:2011fk} and is thus non-universal.

As the strength of interactions is increased, the impurity binds a
particle from the Fermi sea to form a molecule. As in the case of the
polaron state, the molecule is in turn dressed by particle-hole
pairs. The molecule dressed by one particle-hole pair has been studied
in 3D \cite{Combescot2009,Punk:2009kx,Mora:2009qf} and in 2D
\cite{Parish:2011vn} by variational and diagrammatic methods. Here we
extend the diagrammatic method of Combescot {\em et
  al.}~\cite{Combescot2009} to include harmonic confinement. To this
end, we note that the confinement does not change the structure of the
diagrams needed for the molecule energy. The difference in the present
problem is that all fermion propagators are assigned a harmonic
oscillator quantum number and the $T$-matrix depends on
these. Assuming $\ef<\op$ such that in the non-interacting Fermi gas
only the ground state of the harmonic oscillator potential is
occupied, the energy of the dressed molecule is obtained when the
equation
\begin{eqnarray}
  H^{n_1n_2n_3}_{\q\k} & \!\!\!= & \!\!\!\sum_{n_1'n_2'}
\!\! T^{n_1n_2}_{n_1'n_2'}(\q\!-\!\k,\epsilon+\epsilon_{\q0}-\epsilon_{\k
  n_3})
\!\left[1\!-\!n_{F\up}(\k, n_3)\right] \nn \\ 
  &&
  \hspace{-18mm}\times\left\{\sum_{\k'}\frac{H^{n_3n_2'n_1'}_{\q\k'}}{E^{n_1'+n_2'+n_3}_{\k\k'\q}}
    \left[1-n_{F\up}(\k',n_1')\right] 
    -\delta_{0n_1'}\sum_{\q'}\frac{H^{0n_2'n_3}_{\q'\k}}{E^{n_2'+n_3}_{\k}}
  \right. \nn \\ && \left. \hspace{-19mm}
    +\delta_{0n_1'}\!\sum_{n_2''n_3'} \frac{T^{n_2'n_3}_{n_2''n_3'}(\0,\epsilon)}{E_\k^{n_2'+n_3}}
    \sum_{\q'\k'}\frac{H^{0n_2''n_3'}_{\q'\k'}}{E^{n_2''+n_3'}_{\k'}}\left[1-n_{F\up}(\k',n_3')\right]
  \right\}
\label{eq:molecule}
\end{eqnarray}
has a solution.  The vertex $H$ includes all diagrams occurring in
atom-dimer scattering in a quasi-2D geometry \cite{Levinsen2009}.  The
sum on $\q'$ is up to $k_F$ while the sum on $\k'$ is over all
possible momenta. We use the notations
$E_\k^{n}\equiv-\epsilon+\ek+n\op$ and
$E_{\k\k'\q}^{n}\equiv-\epsilon-\eq+\ek+\ek'+\ekkq+n\op$. For further
details see the Supplementary material.

\begin{figure}
\includegraphics[width=\columnwidth]{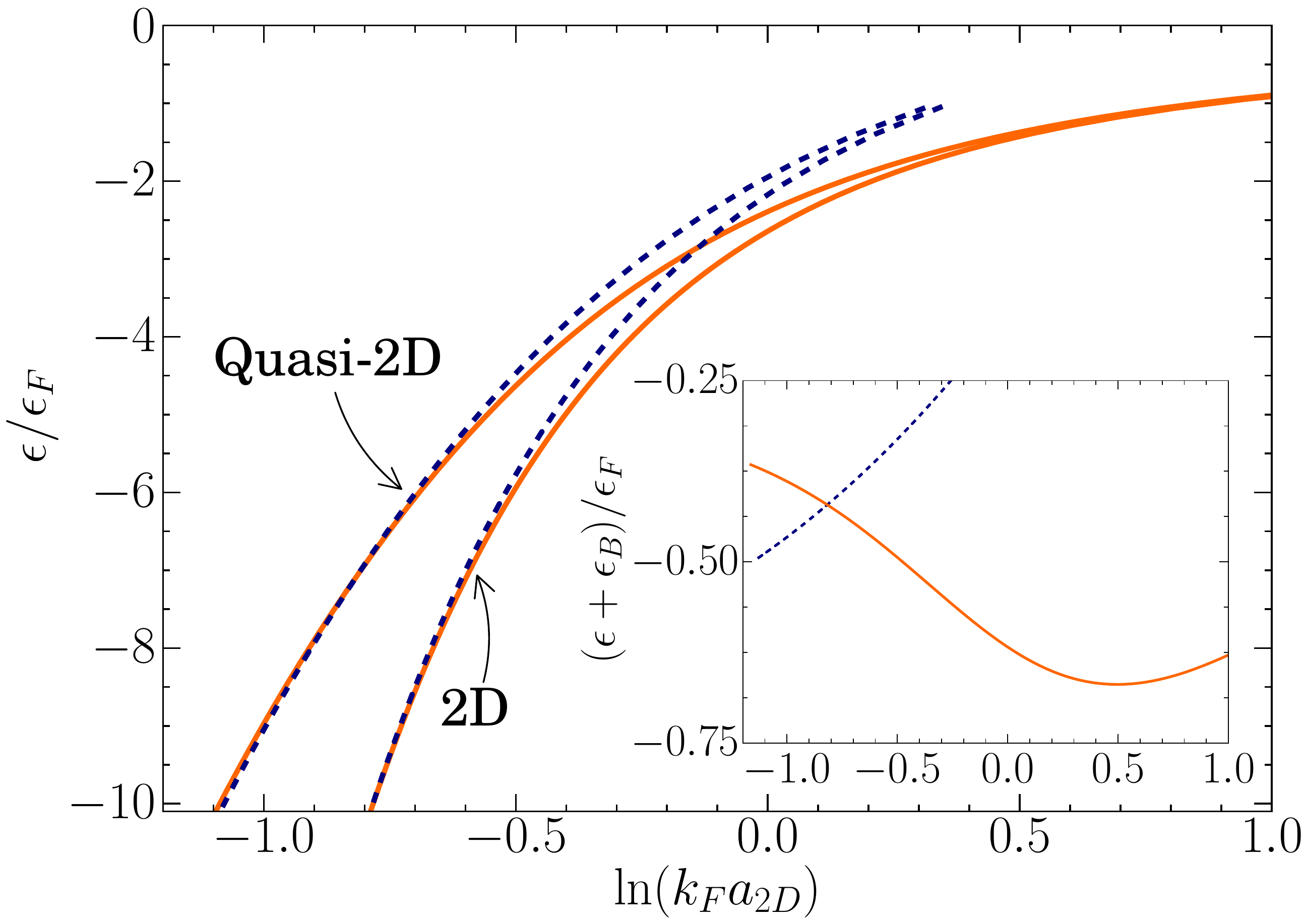}
\caption{(color online). Energy of the attractive polaron (solid line)
  and molecule (dashed) in the 2D regime and in quasi-2D
  ($\ef/\op=1/10$). The molecule energy has been shifted by $\ef$ to
  reflect that a particle has been removed from the Fermi sea. The
  inset shows the quasi-2D energies of the main figure with the
  binding energy subtracted.}
\label{fig:polaronmoleculetransition}
\end{figure}

In Fig. \ref{fig:polaronmoleculetransition} we display the energy of
the attractive polaron and the molecule in the 2D limit and in
quasi-2D. As may be seen, the polaron and molecule energies cross at a
very small angle and thus the position of the polaron-molecule
transition depends sensitively on the confinement.  Computing the
position of the polaron-molecule transition as a function of the
strength of confinement, $\ef/\op$, yields the phase diagram in
Fig. \ref{fig:phasediagram}. As the confinement is weakened from the
2D limit (or the density increased), the transition is seen to move
towards smaller $\eb/\ef$. In particular, in the experiment of
Ref. \cite{Koehl2012} with parameters $\omega_z=2\pi\times78.5$ kHz
and $\ef=2\pi\times10$ kHz the transition was found at $\eb/\ef
\approx 6.8$ while our theory predicts the transition to occur at
$\eb/\ef\approx6.2$ (see Fig. \ref{fig:phasediagram}).  This should be
contrasted with the 2D prediction of $\eb/\ef \approx 10$ at the
polaron molecule transition~\cite{Parish:2011vn}.
 
\begin{figure}
  \includegraphics[width=\columnwidth]{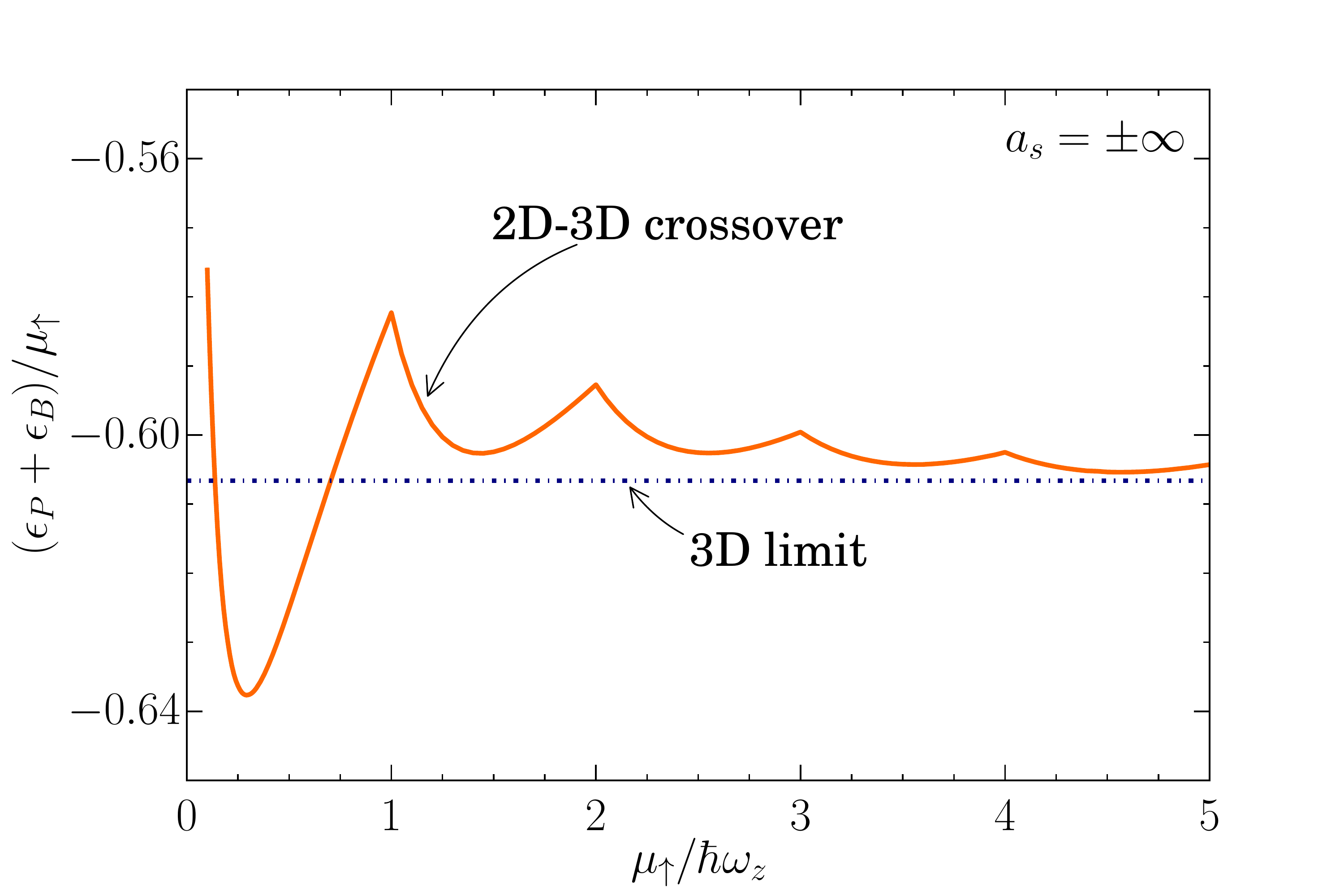}
  \caption{(color online). The energy of the attractive polaron in
    quasi-2D on the 3D Feshbach resonance. The dotted line is the 3D
    limit $\epsilon_P=-0.6066\mu_\up$
    \cite{PhysRevA.74.063628,Combescot:2007bh}.}
\label{fig:dimcross}
\end{figure}

Our formalism is also useful for studying the dimensional crossover
from 2D to 3D in polarized Fermi gases. To illustrate this, we
consider the energy of the attractive polaron on the 3D Feshbach
resonance, where the binding energy of the weakly bound dimer takes
the universal value $\eb=0.244\op$
\cite{RevModPhys.80.885}. Spin-$\up$ atoms in the $n$-th excited
harmonic oscillator level become available whenever the chemical
potential $\mu_\up$ increases through $n\op$ and as a result the
polaron energy is expected to have a discontinuous derivative at each
of these points. Our results are shown in Fig. \ref{fig:dimcross}
where the polaron energy is seen to develop a cusp each time a new
state becomes available due to the increase in the density of $\up$
states. The polaron energy rapidly approaches the 3D limit of an
impurity dressed by one particle-hole pair.

We remark on the resemblance between the above approach and the study
of the ground state of a spin polarized 3D Fermi gas near a narrow
Feshbach resonance
\cite{Massignan2011,Castin2011,Kohstall:2011uq}. The narrowness is due
to a large effective range which enters the problem in addition to the
scattering length and the Fermi wave vector. For fixed scattering
length, an increase in the effective range leads to weaker
interactions and a smaller binding energy. In the quasi-2D problem,
the length-scale $l_z$ plays the role of an effective range: Indeed,
similarly to our results in Fig. \ref{fig:phasediagram} for increased
$l_z$, Refs. \cite{Massignan2011,Castin2011,Kohstall:2011uq} found the
polaron-molecule transition to be shifted towards smaller binding
energies for increased effective range in 3D.

To conclude, we have provided a theory for the highly imbalanced Fermi
gas in quasi-2D. Experimentally, the phase diagram
Fig. \ref{fig:phasediagram} may be mapped out by varying the strength
of the harmonic confinement $\op$ or the density in an ultra-cold
quasi-2D Fermi gas.  Upon entering the molecular phase, the
quasi-particle residue of the impurity abruptly drops to
zero~\cite{Punk:2009kx}, which leaves a clear signature in the
line-shape of RF-spectra~\cite{PhysRevLett.102.230402,
  Koehl2012,Kohstall:2011uq}.  Similarly, the energy of impurities in
ultra-cold Fermi gases have been extracted in quasi-2D from ARPES
spectra, obtained via momentum resolved RF-spectra~\cite{Koehl2012}.

Finally, we point out that while we considered the experimentally more
relevant quasi-2D problem, our results are also of relevance to the
analogous problem of quasi-1D Fermi gases~\cite{Tokatly:2004vn,
  Fuchs:2004kx, Moritz:2005qf, Liao:2010bh}. For example, similar to
our results for the quasi-2D Fermi gas, one expects that the phase
diagram of the 1D imbalanced gas provides an inadequate description at
stronger coupling. 

\begin{acknowledgments}
  We gratefully acknowledge insightful discussions with M.~M.~Parish,
  N. ~R.~Cooper, A.~M.~Fischer and M.~K\"ohl. We thank the group of M. K\"ohl for kindly sharing their unpublished data. JL acknowledges support
  from a Marie Curie Intra European grant within the 7th European
  Community Framework Programme. SKB acknowledges support from the
  EPSRC (Grant EP/I010580/1).
\end{acknowledgments}

\bibliography{quasi2d_skb,polaron,quasi2dextra}

\begin{thebibliography}{10}

\bibitem{Frohlich:2011cr}
B. Fr\"ohlich {\it et~al.}, Phys. Rev. Lett. {\bf 106},  105301  (2011).

\bibitem{Martiyanov:2010kl}
K. Martiyanov, V. Makhalov, and A. Turlapov, Phys. Rev. Lett. {\bf 105},
  030404  (2010).

\bibitem{Dyke:2011tg}
P. Dyke {\it et~al.}, Phys. Rev. Lett. {\bf 106},  105304  (2011).

\bibitem{Feld:2011nx}
M. Feld {\it et~al.}, Nature {\bf 480},  75  (2011).

\bibitem{sommer2011_2D}
A.~T. Sommer {\it et~al.}, Phys. Rev. Lett. {\bf 108},  045302  (2012).

\bibitem{Zhang:2012uq}
Y. {Zhang}, W. {Ong}, I. {Arakelyan}, and J.~E. {Thomas}, Phys. Rev. Lett. {\bf
  108},  235302  (2012).

\bibitem{Koehl2012}
M. Koschorreck {\em et al.}, Nature {\bf 485}, 619 (2012); M. K\"ohl, talk at
  the 2012 APS March meeting.

\bibitem{Parish:2011vn}
M.~M. Parish, Phys. Rev. A {\bf 83},  051603  (2011).

\bibitem{Petrov:2001fk}
D.~S. Petrov and G.~V. Shlyapnikov, Phys. Rev. A {\bf 64},  012706  (2001).

\bibitem{PhysRevB.77.020408}
N. Prokof'ev and B. Svistunov, Phys. Rev. B {\bf 77},  020408  (2008).

\bibitem{PhysRevA.74.063628}
F. Chevy, Phys. Rev. A {\bf 74},  063628  (2006).

\bibitem{PhysRevLett.102.230402}
A. Schirotzek, C.-H. Wu, A. Sommer, and M.~W. Zwierlein, Phys. Rev. Lett. {\bf
  102},  230402  (2009).

\bibitem{schmidt2011}
R. Schmidt, T. Enss, V. Pietil\"a, and E. Demler, Phys. Rev. A {\bf 85},
  021602  (2012).

\bibitem{Ngampruetikorn:2011fk}
V. {Ngampruetikorn}, J. {Levinsen}, and M.~M. {Parish}, Europhys. Lett. {\bf
  98},  30005  (2012).

\bibitem{LL}
L.~D. Landau and E.~M. Lifshitz, {\em Quantum Mechanics}
  (Butterworth-Heinemann, Oxford, UK, 1981).

\bibitem{RevModPhys.80.885}
I. Bloch, J. Dalibard, and W. Zwerger, Rev. Mod. Phys. {\bf 80},  885  (2008).

\bibitem{Pietila:2011hc}
V. Pietil\"a, D. Pekker, Y. Nishida, and E. Demler, Phys. Rev. A {\bf 85},
  023621  (2012).

\bibitem{Combescot:2007bh}
R. Combescot, A. Recati, C. Lobo, and F. Chevy, Phys. Rev. Lett. {\bf 98},
  180402  (2007).

\bibitem{Zollner:2011fk}
S. Z\"{o}llner, G.~M. Bruun, and C.~J. Pethick, Phys. Rev. A {\bf 83},
  021603(R)  (2011).

\bibitem{Klawunn2011}
M. Klawunn and A. Recati, Phys. Rev. A {\bf 84},  033607  (2011).

\bibitem{Kohstall:2011uq}
C. {Kohstall} {\it et~al.}, Nature {\bf 485},  615  (2012).

\bibitem{PhysRevLett.101.050404}
R. Combescot and S. Giraud, Phys. Rev. Lett. {\bf 101},  050404  (2008).

\bibitem{Massignan:2011fk}
P. Massignan and G.~M. Bruun, Eur. Phys. J. D {\bf 65},  83  (2011).

\bibitem{Combescot2009}
R. Combescot, S. Giraud, and X. Leyronas, Europhys. Lett. {\bf 88},  60007
  (2009).

\bibitem{Punk:2009kx}
M. Punk, P. Dumitrescu, and W. Zwerger, Phys. Rev. A {\bf 80},  053605  (2009).

\bibitem{Mora:2009qf}
C. Mora and F. Chevy, Phys. Rev. A {\bf 80},  033607  (2009).

\bibitem{Levinsen2009}
J. Levinsen, T. Tiecke, J. Walraven, and D. Petrov, Phys. Rev. Lett. {\bf 103},
   153202  (2009).

\bibitem{Massignan2011}
P. {Massignan}, Europhys. Lett. {\bf 98},  10012  (2012).

\bibitem{Castin2011}
C. {Trefzger} and Y. {Castin}, \pra {\bf 85},  053612  (2012).

\bibitem{Tokatly:2004vn}
I.~V. Tokatly, Phys. Rev. Lett. {\bf 93},  090405  (2004).

\bibitem{Fuchs:2004kx}
J.~N. Fuchs, A. Recati, and W. Zwerger, Phys. Rev. Lett. {\bf 93},  090408
  (2004).

\bibitem{Moritz:2005qf}
H. Moritz {\it et~al.}, Phys. Rev. Lett. {\bf 94},  210401  (2005).

\bibitem{Liao:2010bh}
Y.-a. Liao {\it et~al.}, Nature {\bf 467},  567  (2010).

\bibitem{chasman1967}
R.~R. Chasman and S. Wahlborn, Nucl. Phys. A {\bf 90},  401  (1967).

\end{thebibliography}


\newpage

\begin{widetext}

\newpage

\begin{center}
{\large \bf Supplementary material} \\
(Dated: \today)
\vspace{5mm}
\end{center}

\begin{center}
Jesper Levinsen and Stefan K. Baur \\
{\it Cavendish Laboratory, JJ Thomson Avenue, Cambridge, CB3 0HE, United Kingdom} 
\end{center}

\section*{$T$-matrix in the medium}

Consider the scattering of two particles at 2D momentum $\q$ and
energy $\epsilon$. The $T$-matrix describing forward scattering in the
medium in the presence of harmonic confinement was recently calculated
in Ref. \cite{Pietila:2011hc}, we include it here for completeness:
For incoming and outgoing harmonic oscillator quantum numbers
$n_1,\,n_2$ and $n_1',\,n_2'$, the $T$-matrix describing
forward scattering in the medium takes the form
\begin{equation}
  T^{n_1n_2}_{n_1'n_2'}(\q,\epsilon)=\sqrt{2\pi}l_z\sum_{NN'n_rn_r'}C^{n_1n_2}_{Nn_r}
C^{n_1'n_2'}_{N'n_r'}\phi_{n_r}^*(0) \phi_{n_r'}(0)\T_{N,N'}(\q,\epsilon),
\end{equation} 
in terms of relative and center of mass quantum numbers $n_r,\,n_r'$ and
$N,\,N'$. The Clebsch-Gordan coefficients for the change of basis are
given by \cite{chasman1967}
\begin{equation}
C^{n_1n_2}_{Nn_r}=\delta_{n_1+n_2,N+n_r} 2^{-\frac12(n_1+n_2)}\sqrt{\frac{N!n_r!}{n_1!n_2!}}
\sum_{m=\max(n_r-n_1,0)}^{\min(n_r,n_2)}(-1)^m
\left(\begin{array}{cc}n_1\\n_r-m\end{array}\right)
\left(\begin{array}{cc}n_2\\m\end{array}\right),
\end{equation}
and the harmonic oscillator wavefunction $\phi$ takes
the value
\begin{equation}
\sqrt{2\pi}l_z\left|\phi_n(0)\right|^2=\left\{\begin{array}{ll}\frac{(n-1)!!}{n!!},
    & $n$ \mbox{ even} \\
0 & $n$ \mbox{ odd}\end{array}
\right.
\end{equation}
at the origin. $\T_{NN'}$ is renormalized by the use of the two-body
$T$-matrix as \cite{Pietila:2011hc}
\begin{equation}
\T_{NN'}^{-1}(\q,\epsilon)=\T_0^{-1}(\epsilon-N\op-\eq/2)\delta_{NN'}-
{\cal D}_{NN'}(\q,\epsilon),
\end{equation}
with the renormalized polarization operator given by
\begin{equation}
{\cal
  D}_{NN'}(\q,\epsilon)=-\sum_{n_1n_2\k}u_{n_1+n_2-N,n_1+n_2-N'}C^{n_1n_2}_{N',n_1+n_2-N'}C^{n_1n_2}_{N,n_1+n_2-N}
\frac{n_{F\up}(\q-\k,n_1)+n_{F \down }(\k,n_2)}{\epsilon-\ekq-\ek-\op(n_1+n_2)} 
\end{equation}
and expansion coefficients
$u_{n,m}=(-1)^{(n+m)/2}(n-1)!!(m-1)!!/\sqrt{n!m!}$ for $n,m$ even
non-negative integers, 0 otherwise. The Fermi function
$n_{F\up}(\k,n_1)$ is 1 for a particle in the $\up$ Fermi sea, 0
otherwise. For the polarized gas considered in this Letter, the Fermi
function $n_{F\down}(\k,n_2)$ vanishes.

\section*{Equation for the dressed molecule}

We now construct the sum of diagrams needed to obtain the energy of
the molecular state dressed by one particle-hole pair in the presence
of harmonic confinement. Formally, the diagrams are identical to those
considered in Ref. \cite{Combescot2009} for the 3D problem, although
in the present problem all fermion lines have additional harmonic
oscillator quantum numbers. For simplicity, we restrict ourselves to
a molecule at rest. We also require $\ef<\op$ such that hole
propagation always proceeds in the lowest harmonic oscillator
state. Our results are easily generalizable to finite momentum and
occupation of higher bands.

The bound state energy corresponds to a divergence of a two-particle
vertex. As argued in Ref. \cite{Combescot2009}, the energy of the
molecule dressed by one particle-hole pair necessarily lies below the
energy of the bare molecule. Thus it suffices to consider the vertex
$H$ describing the sum of diagrams containing a maximum of two forward
propagating $\up$ atoms as well as the impurity atom.  The integral
equation satisfied by $H$ is depicted in Fig. \ref{fig:H}: In the
vertex $H$ the initial interaction is between the impurity and the
$\up$ atom with quantum number $n_1$. After this interaction, this
particle either closes its own loop (the last two diagrams on the
r.h.s. of Fig. \ref{fig:H}) or participates in another vertex $H$.
Taking into account only the homogenous terms (any pole in the
non-homogenous term lies above the dressed molecule energy) finally
results in the integral equation (\ref{eq:molecule}) in the main text.

Additionally, we have verified that a variational approach with the
trial wave functions for the polaron
\begin{eqnarray}
\ket{P}=\sum_n \phi_n c_{\down \mathbf{0}n}^{\dagger} \ket{\text{FS},
  N_{\up}}+\sum_{\substack{ nn'm \\ \mathbf{k} \mathbf{q}}} 
\phi^{nn'm}_{\mathbf{k}\mathbf{q}} c_{\down \mathbf{q}-\mathbf{k}
  n}^{\dagger}
c_{\up n' \mathbf{k}}^{\dagger} c_{\up m \mathbf{q}} \ket{\text{FS}, N_{\up}}
\end{eqnarray}
and the molecule
\begin{eqnarray}
\ket{M}=\sum_{nn'\mathbf{k}}\phi_k^{nn'} c_{\up \mathbf{k}n}^{\dagger} 
c_{\down-\mathbf{k}n'}^{\dagger} \ket{\text{FS}, N_{\up}-1}
+\frac{1}{2} 
\sum_{\substack{nn'mm' \\ \mathbf{k}\mathbf{k}'\mathbf{q}}} 
\phi^{nn'mm'}_{\mathbf{k}\mathbf{k}'\mathbf{q}} 
c_{\down \mathbf{q-k-k'}n}^{\dagger}
c_{\uparrow \mathbf{k} n'}^{\dagger}
c_{\uparrow \mathbf{k}' m}^{\dagger}
c_{\uparrow \mathbf{q} m'}\ket{\text{FS}, N_{\up}-1}
\end{eqnarray}
yields the equations (\ref{eq:polaron}) for the polaron and
(\ref{eq:molecule}) for the molecule quoted in the main text. Here
$c_{\sigma \mathbf{k} n}^{\dagger}$ denotes the creation operator for
a particle of spin $\sigma$ with in-plane momentum $\mathbf{k}$ in the
n-th transverse harmonic oscillator mode. $\ket{\text{FS}, N_{\up}}$
is the non-interacting ground-state of $N_{\up}$ spin-$\up$ fermions.

\begin{figure}
\includegraphics[width=\columnwidth]{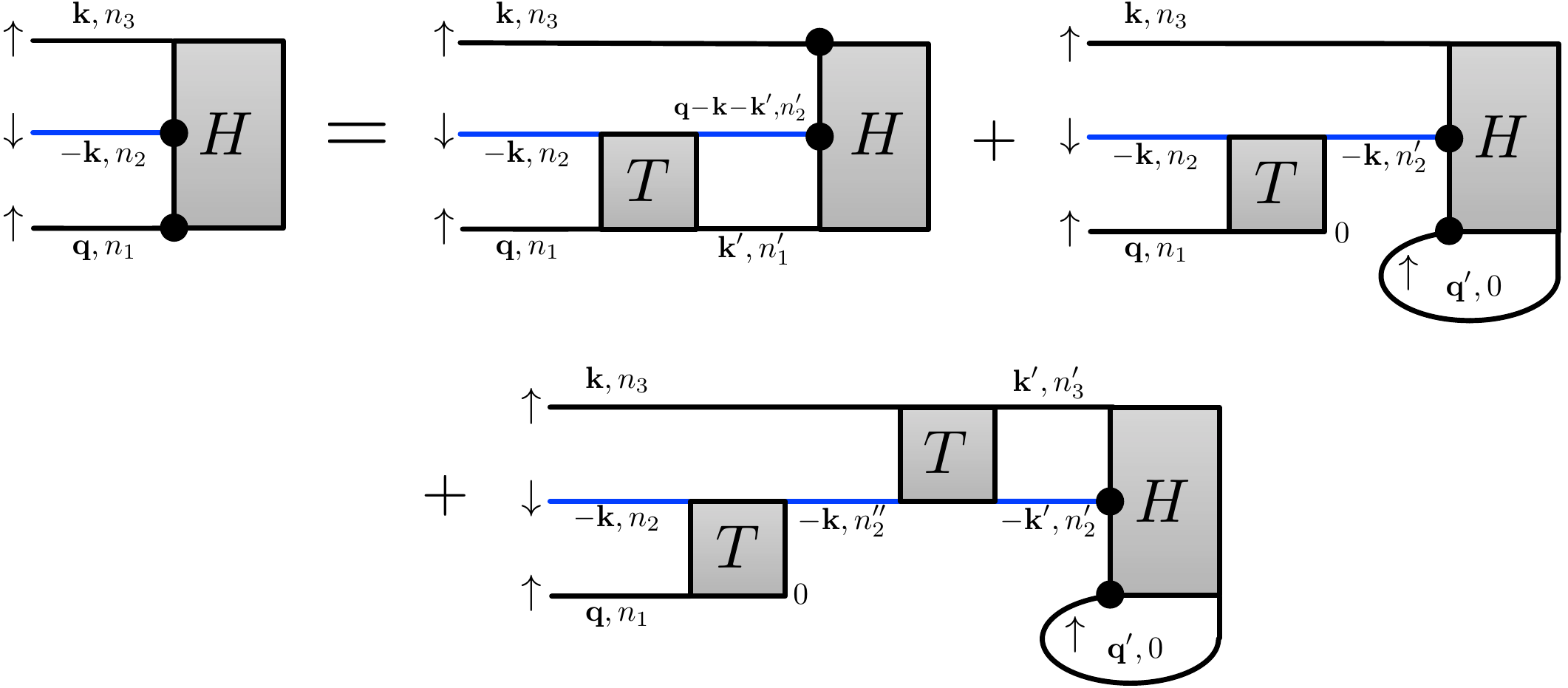}
\caption{(color online). The integral equation (\ref{eq:molecule})
  satisfied by the vertex $H$.  Circles indicate that the interaction
  inside the vertex is initially between the two marked particles. The
  loop on the last two diagrams yields a summation over $\q'$.}
\label{fig:H}
\end{figure}

\newpage

\end{widetext}


\end{document}